\documentclass[aps,twocolumn,prb,amsmath,amssymb,showpacs,superscriptaddress,floatfix]{revtex4-1}

\usepackage{color,soul}
\usepackage{graphicx,epsfig}
\usepackage[hypertex]{hyperref}

\DeclareMathOperator{\Cs}{C_{\scriptscriptstyle{\Sigma}}}
\DeclareMathOperator{\Li}{Li}\DeclareMathOperator{\Brnli}{\mathrm B}
\DeclareMathOperator{\ssL}{\scriptscriptstyle{\mathrm{L}}}
\DeclareMathOperator{\ssR}{\scriptscriptstyle{\mathrm{R}}}
\DeclareMathOperator{\LR}{\scriptscriptstyle{\mathrm{(LR)}}}
\DeclareMathOperator{\RL}{\scriptscriptstyle{\mathrm{(RL)}}}

\begin{document}
\date{\today }

\title{Interplay of charge and heat transport in a nano-junction in the out-of-equilibrium cotunneling regime}

\author{N.~M.~Chtchelkatchev}
\affiliation{Institute for High Pressure Physics, Russian Academy of Science, Troitsk 142190, Russia}
\affiliation{Department of Theoretical Physics, Moscow Institute of Physics and Technology, 141700 Moscow, Russia}

\author{A.~Glatz}
\affiliation{Materials Science Division, Argonne National Laboratory, Argonne, Illinois 60439, USA}
\affiliation{Department of Physics, Northern Illinois University, DeKalb, Illinois 60115, USA}

\author{I.~S.~Beloborodov}
\affiliation{Department of Physics and Astronomy, California State University Northridge, Northridge, CA 91330, USA}

\begin{abstract}
We study the charge transport and the heat transfer through a nano-junction composed of a small metallic grain weakly coupled to two metallic leads.
We focus on the cotunneling regime out-of-equilibrium, where the bias voltage and the temperature gradient between the leads strongly drive electron and phonon degrees of freedom in the grain that in turn have a strong feedback on the transport through the grain.
We derive and solve coupled kinetic equations for electron and phonon degrees of freedom in the grain.
We obtain the heat fluxes between cotunneling electrons, bosonic electron-hole excitations in the grain, and phonons, and self-consistently find the current-voltage characteristics.
We demonstrate that the transport in the nano-junction is very sensitive to the spectrum of the bosonic modes in the grain.
\end{abstract}

\maketitle

\section{Introduction}

Quantum composite materials are the subject of current intensive research activity. A typical quantum  composite material (QCM) is a granular conductor, where the grains are so small that the typical charging energy of one electron, $E_c$, is the largest energy scale, in particular it is larger than temperature and voltage.

The equilibrium properties of granular conductors are well understood.~\cite{Beloborodov07}
However, much less is known about the transport properties of quantum composite conductors out-of-equilibrium, e.g., at
large bias or at large enough temperature gradients, when linear response theory is not valid.

There are several transport channels for electron propagation though the granular QCM.
At weak coupling between the grains and not very high temperatures cotunneling is the main mechanism of electron transport.~\cite{Beloborodov05}
It provides a conduction channel at low applied biases and temperatures, where otherwise the Coulomb blockade arising from the electron-electron repulsion would suppress the current flow.
The essence of a cotunneling process is that an electron tunnels via virtual states thus bypassing the large Coulomb barrier.~\cite{Averin}
Here we focus on the out-of-equilibrium cotunneling through a single grain, see Fig.~\ref{fig.junction}a), the building block of a quantum composite material, attached to two bulk leads.
Voltage and/or temperature differences between the leads strongly drive electron and phonon degrees of freedom in the grain.
The solution of the transport problem implies a selfconsistent calculation, which takes into account the mutual feedback of highly excited degrees of freedom in the grain on the cotunneling electrons and vice-versa.

Inelastic cotunneling of electrons from one lead to the other through a grain is accompanied by the creation of electron-hole pairs in the granule that results through its decay in heating of the grain.
On the other hand, electrons in the grain exchange energy with phonons.
Thus, the electron temperature in the granule depends on the balance between heating by electron-hole pairs and cooling by phonons.
Recently heating effects in a single grain and a chain of grains due to inelastic cotunneling were discussed in several papers.~\cite{GlatzPRB10,glatz+prb10,Nazarov_new,Nazarov_PRB}
However, our consideration was in the linear response regime, i.e., valid for small gradients of voltage and temperature.
Here we follow a general approach, which also holds in non-equilibrium situations and, furthermore, derive our results from microscopic considerations.

In the following sections, we first present our model and main results, followed by their derivations, and then discuss our results in detail.

\section{Model and main results}

\begin{figure}[tbh]
  \includegraphics[width=0.4\textwidth]{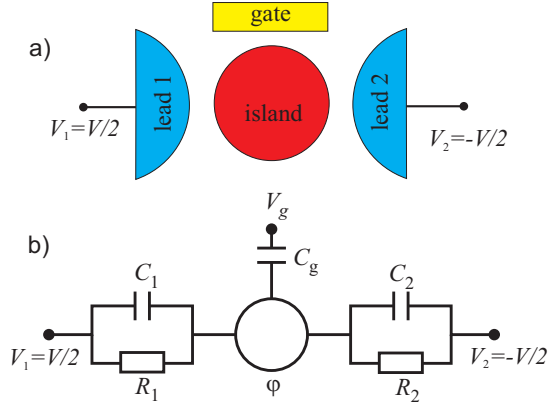}\\
  \caption{(Color online) Sketch of a nano-junction, a), and its equivalent circuit scheme, b). }\label{fig.junction}
\end{figure}

We investigate out-of-equilibrium properties of a nano-junction shown in Fig.~\ref{fig.junction}a).
The single grain in the junction is characterized by two energy scales: (i) the mean energy level spacing $\delta$ 
and (ii) the charging energy $E_c$.~\cite{Beloborodov07,Aleiner}
Here we are interested in the case of metallic grains, where $E_c \gg \delta$.
In addition to theses two energy scales the system is characterized by the bare tunneling resistance $R_{\ssL(\ssR)}$ between the left (right) lead and the grain.
In this paper we concentrate on the Coulomb blockade regime, corresponding to a weak coupling between the grain and the leads, $R_{\ssL(\ssR)} \gg R_{\rm q}$ with $R_{\rm q} = \pi \hbar/e^2$ being the quantum resistance, Fig.~\ref{fig.junction}.
Since we focus on the cotunneling regime in this work, our considerations are valid for temperatures $\delta < T < E_c$ and the leads are assumed to be heat sinks and stay at ambient temperatures $T_{\ssL, \ssR}$, since they are much larger than the grain.

As a main result, it turns out that the transport characteristics of these junctions are very sensitive to the temperature of the electron degrees of freedom in the grain, $T_g$.  The grain temperature, $T_g(V,T_{\ssL},T_{\ssR})$, is found explicitly as a function of voltage bias $V$ and both lead temperatures from the expression for the heat fluxes in the nano-junction out-of-equilibrium. Finally, this temperature is used to obtain the current-voltage characteristics, $I(V)$.

The heat, $\dot Q$, generated in the grain due to inelastic electron cotunneling can be derived as follows -- representing the first main result of our work:
\begin{gather}\label{Q}
  \dot Q(T_{\ssL},T_{\ssR},V) =a\, [T_{\rm eff}^4-T_g^4],
  \\
 T_{\rm eff}^4=T_{\rm m}^4+\frac5{2\pi^2}T_{\rm m}^2(eV)^2+\frac{5}{8\pi^4}(eV)^4,
\end{gather}
where $T_{\rm m}^2=\left(T_{\ssL}^2+T_{\ssR}^2\right)/2$ and a material dependent constant $a$, which is derived below [see Eq.~(\ref{a1})].
The energy scale $T_{\rm eff}$ has the physical meaning of an effective temperature of the cotunneling electrons.
It is important to note that the r.~h.~s. in Eq.~\eqref{Q} can be positive or negative indicating a heating or cooling of the grain, respectively. 
For equal temperatures $T_{\ssL}=T_{\ssR}$ and zero voltage $V=0$ we reproduce the results of Ref.~[\onlinecite{kravtsov}] from Eq.~\eqref{Q}.

The heat dissipation in the granule goes through two stages.
First, electron-hole pairs are excited by inelastic cotunneling processes and second the electron-hole pairs recombine and release their excess energy to the phonon bath.
The heat flux between phonons and electrons (electron-hole pairs) has the form~\cite{RMP06,Altshuler09}
\begin{gather}\label{ephq}
  \dot q(T_g,T_{\rm ph}) = (T_g^\alpha-T_{\rm ph}^\alpha)\,\kappa,
\end{gather}
where $T_{\rm ph}$ is the temperature of the phonon bath and the parameters $\kappa$ and $\alpha$  depend on the particular model of electron-phonon interaction; typically  $\alpha =4,5$ or $6$.

In order to obtain the grain temperature, we have to solve the heat balance equation in the grain
\begin{gather}\label{heatballance}
  \dot q(T_g,T_{\rm ph}) = \dot Q(T_{\ssL},T_{\ssR},V),
\end{gather}
where  $\dot Q(T_{\ssL},T_{\ssR},V)$ is given in Eq.~(\ref{Q}).

It is useful to express all temperatures\cite{units} and voltages (more precisely $eV$) in terms of the charging energy $E_c$, which we will denote by a tilde, e.g., $\tilde T_{\rm eff}=T_{\rm eff}/E_c$ or $\tilde V=eV/E_c$.

\begin{figure}[t]
\begin{center}
  \includegraphics[width=0.95\columnwidth]{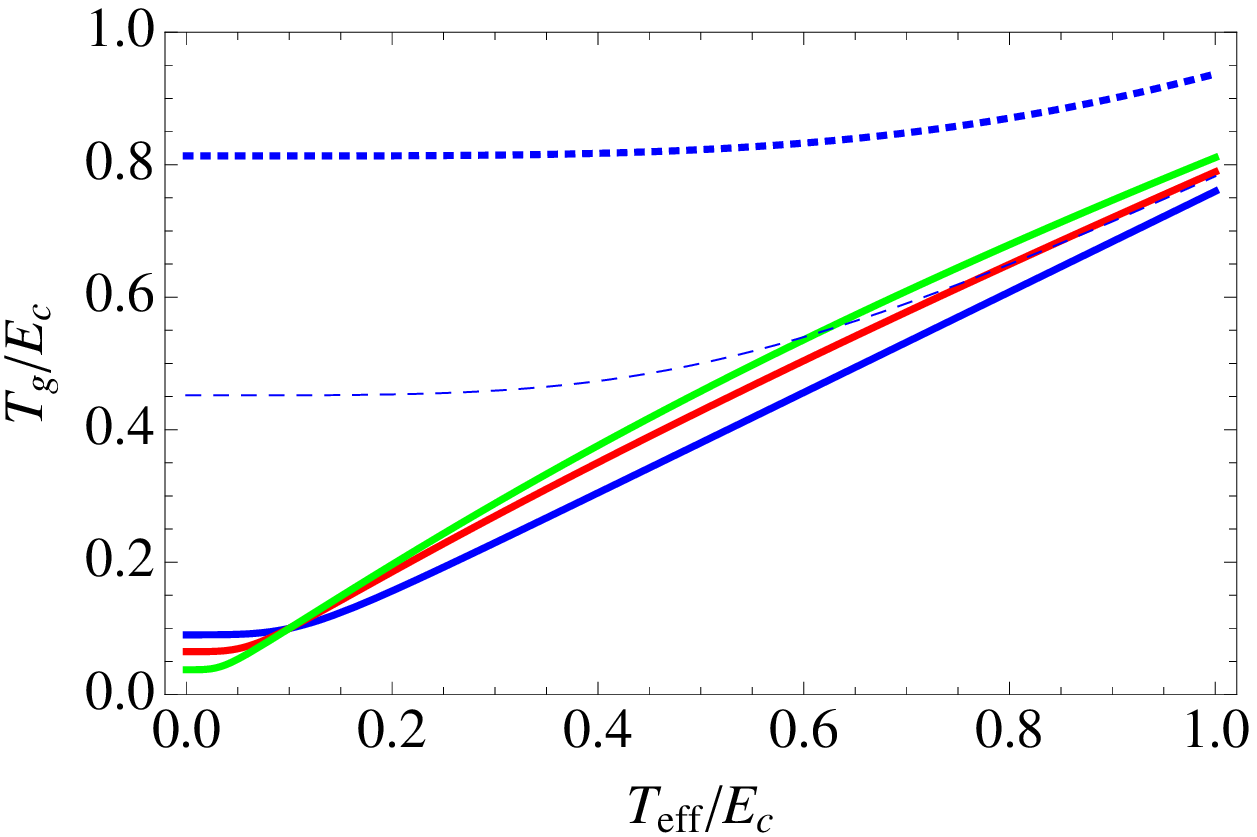}\\
  \includegraphics[width=0.95\columnwidth]{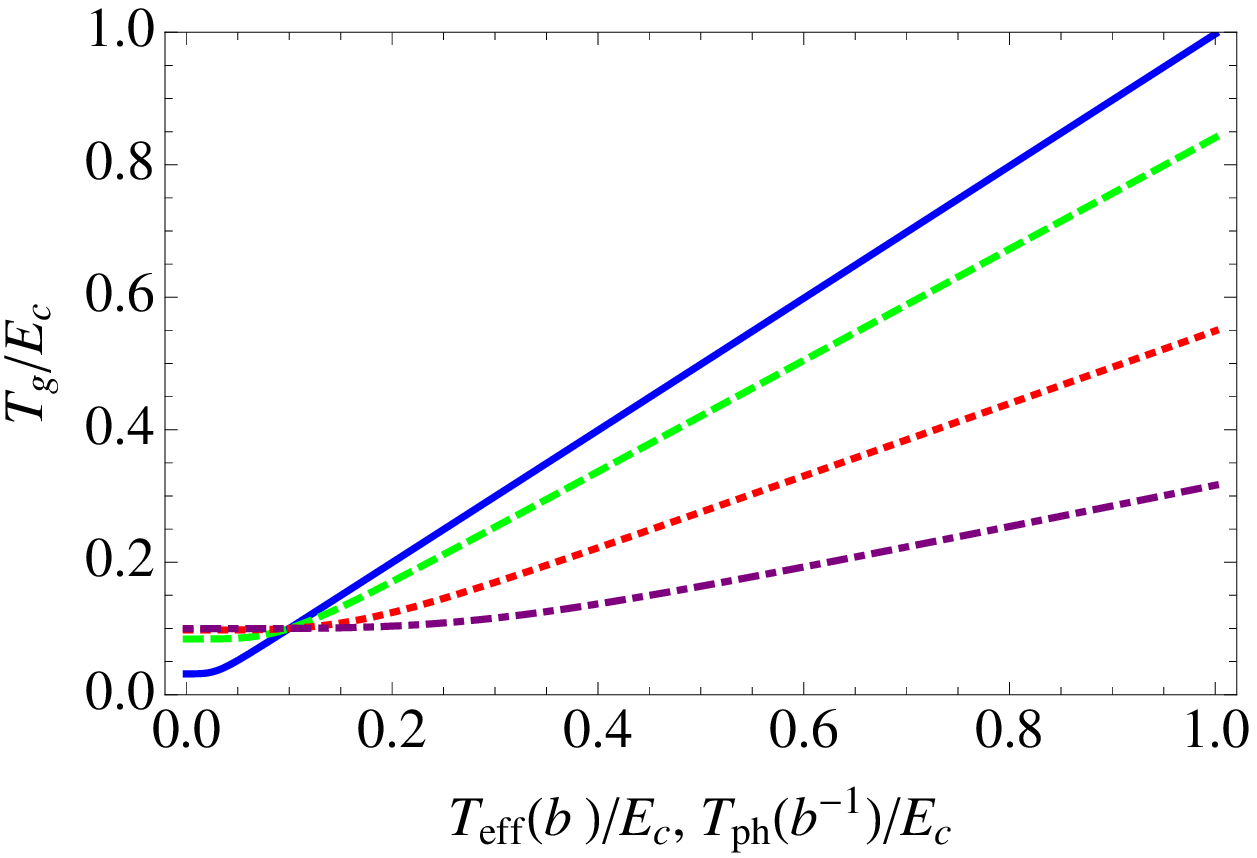}
  \end{center}
  \caption{(Color online) \textit{(top)} Behavior of the grain temperature $T_g$ as functions of $T_{\rm eff}$. The blue curves are for $\alpha=4$, red for $\alpha=5$, and green for $\alpha=6$. The solid line are for $\tilde T_{\rm ph}=0.1$, the dashed line for  $\tilde T_{\rm ph}=0.5$, and the dotted thick lines for $\tilde T_{\rm ph}=0.9$.  In all curves the dimensionless parameter is fixed to $b=0.5$ (see text). 
  \textit{(bottom)} Behavior of $T_g$ as functions of $T_{\rm eff}$ for $\alpha=4$, $\tilde T_{\rm ph}=0.1$, and different $b$. The solid curve is for $b=0.01$, dashed for $b=1$, dotted for $b=10$, and dash-dotted for $b=100$.  The dependence of $T_g$ on $T_{\rm ph}$ for fixed $T_{\rm eff}$ looks similar if $b$ is replaced by $b^{-1}$ (exactly the same for $\alpha=4$).}\label{fig.Tg}
\end{figure}

Introducing the dimensionless parameter $b=aE_c^{4-\alpha}/\kappa$, Eq.~(\ref{heatballance}) can be written as
\begin{equation}\label{heatballance2}
 \tilde T_g^\alpha-\tilde T_{\rm ph}^\alpha=b\left(\tilde T_{\rm eff}^4- \tilde T_g^4\right)\,.
\end{equation}

For $\alpha=4$, Eq.~\eqref{heatballance2} can be solved analytically:
\begin{gather}\label{Tg_alpha4}
  \tilde T_g^4= \left(T_{\rm ph}^4+bT_{\rm eff}^4\right)/(1+b)\,.
\end{gather}

The behavior of the grain temperature $T_g$ as a function of $T_{\rm eff}$ for different $\alpha$-values and values of 
parameter $b = 0.01, 1, 10, 100$ is presented in Fig.~\ref{fig.Tg}.

Using Eq.~\eqref{Tg_alpha4} for the grain temperature $T_g$, we can finally calculate the current-voltage characteristics due to inelastic cotunneling
\begin{equation}
\label{IV1}
I^{\rm (inel)}(\tilde V) = \frac{1}{12\pi^3} \frac{R_q^2}{R_{\ssL}R_{\ssR}} \frac{eE_c}{\hbar} 
\left\{\tilde V^2+2\pi^2(\tilde T_{\rm m}^2+\tilde T_g^2)\right\}\tilde V\,,
\end{equation}
which is the second main results of this paper.

When deriving Eqs.~\eqref{Q}-\eqref{Tg_alpha4}, we neglect possible sequential electron tunneling.
The latter is suppressed by the Coulomb blockade if temperature and voltage are below the characteristic single-electron charging energy in the grain, $eV, T < E_c$ with $E_c\sim e^2/\Cs$, where $\Cs=C_{\ssL}+C_{\ssR}+C_g$ is the total capacitance, see Fig.~\ref{fig.junction}b). Furthermore, the leads are assumed to be in equilibrium at similar temperatures and energy relaxation processes within the grain are fast such that we can use a local equilibrium description.

In the following section we present the general out-of-equilibrium description of the heat and charge transport through the junction, which yields the derivation of  Eqs.~\eqref{Q}-\eqref{Tg_alpha4} and eventually results in Eq.~(\ref{IV1}).

\section{Cotunneling transport}

We start with the most general expression for the current-voltage characteristics due to inelastic electron cotunneling.

\subsection{Inelastic cotunneling rate}
The inelastic cotunneling rate for an electron transfer from the left to the right lead has the following form:~\cite{Averin92}

\begin{widetext}
\begin{multline}\label{inelastic_rate}
 \overrightarrow{\Gamma}^{\rm(inel)} =  \frac{1}{2\pi^3} \frac{R_q^2}{R_{\ssL}R_{\ssR}} \frac{1}{\hbar} \int d\epsilon_1d\epsilon_2d\epsilon_3d\epsilon_4f^{(L)}(\epsilon_1)[1-f^{(g)}(\epsilon_2)]f^{(g)}(\epsilon_3)[1-f^{(R)}(\epsilon_4)]\times
 \\
 \left\{\frac{1}{\epsilon_2-\epsilon_1+\Delta F_{\ssL}^+(n)}+\frac{1}{\epsilon_4-\epsilon_3+\Delta F_{\ssR}^-(n)}\right\}^2
 \delta(eV+\epsilon_1-\epsilon_2+\epsilon_3-\epsilon_4),
\end{multline}
\end{widetext}
where $V$ is the bias voltage between the left and right leads.
The distribution functions $f^{(\ssL,\ssR)}(\epsilon)$ describe electrons within the left and right leads, respectively.

If the temperature of the left (right) lead is $T_{\ssL(\ssR)}$ then $f_{\ssL(\ssR)}(\epsilon)=f_F(\epsilon,T_{\ssL(\ssR)})$, with $f_F(\epsilon,T)=1/[\exp(\epsilon/T)+1]$ being the Fermi-Dirac distribution function.
Function $f^{(g)}(\epsilon)$ in Eq.~\eqref{inelastic_rate} is the electron distribution in the grain.
The functions $\Delta F_{\ssL(\ssR)}^+(n)$ denote the change of the charging energy when an electron tunnels from the left (right) lead onto the island with  excess charge $n e$, while $\Delta F_{\ssL(\ssR)}^-(n)$ denotes the change of the charging energy when an electron tunnels out of the island with  excess charge $n e$ to the left (right) lead
\begin{subequations}
\begin{gather}\label{eq:dF1}
    \Delta F_{\ssL}^\pm=\frac{e^2}{\Cs}\left\{\left[\frac12\pm\left(n+\frac{C_gV_g}e\right)\right]\mp
    \frac{[C_{\ssR}+C_g/2]V}e\right\},
    \\\label{eq:dF2}
    \Delta F_{\ssR}^\pm=\frac{e^2}{\Cs}\left\{\left[\frac12\pm\left(n+\frac{C_gV_g}e\right)\right]\pm
    \frac{[C_{\ssL}+C_g/2]V}e\right\},
\end{gather}
\end{subequations}
where all the capacitances are defined in Fig.~\eqref{fig.junction}b) and $\Cs$ is the total capacitance introduced below Eq.~\eqref{IV1}.
The number of excess electrons on the grain is determined by the condition that $n$ results in the minimal electrostatic energy: $\Delta F_{\ssL}^+(n)>0$ and $\Delta F_{\ssR}^-(n)>0$,~\cite{Averin92}.

The inelastic electron cotunneling rate $\overleftarrow{\Gamma}^{\rm(inel)}$ from the right to the left lead
can be found using Eq.~\eqref{inelastic_rate} with proper substitutions: $V\to-V$ in the $\delta$-function
and interchange between the left and the right leads, ``L'' $\leftrightarrows$ ``R''.

\subsection{Cotunneling current}

To write the total cotunneling current we also take into account elastic electron cotunneling rates, $\overrightarrow{\Gamma}^{\rm(el)}$ and $\overleftarrow{\Gamma}^{\rm(el)}$, which can be written similarly to Eq.~\eqref{inelastic_rate}, i.e.,
\begin{gather}\label{current-voltage}
I(V)=e\left(\overrightarrow{\Gamma}^{\rm(inel)}-\overleftarrow{\Gamma}^{\rm(inel)}+\overrightarrow{\Gamma}^{\rm(el)}-\overleftarrow{\Gamma}^{\rm(el)}\right).
\end{gather}
Here we stress that the rates and the current in Eq.~\eqref{current-voltage} strongly depend on the electron distribution function $f^{(g)}(\epsilon)$ in the grain. This distribution can be approximated by an equilibrium distribution with phonon bath temperature, $T_{\rm ph}$, only for small voltage and temperature gradients across the junction.~\cite{Averin92} In this paper we consider the current-voltage characteristics beyond the equilibrium approximation. We write the kinetic equation governing the behavior of function $f^{(g)}(\epsilon)$ and find the dependence of grain temperature $T_g$ on the bias voltage $V$, lead temperatures $T_{\ssL}$, $T_{\ssR}$, and the phonon temperature, $T_{\rm ph}$.

In the following we concentrate on the inelastic component of the cotunneling current $I^{(\rm inel)}(V)$ which heats the grain in contrast 
to the elastic part.

\section{Kinetic equation for $f^{(g)}$}

To calculate the current-voltage characteristics, $I(V)$, in Eq.~\eqref{current-voltage} we need to know the electron distribution
function $f^{(g)}(\epsilon)$ in the grain. This distribution function satisfies the kinetic equation
\begin{equation}\label{keq}
 \frac{d}{d (\delta t)} f^{(g)}(\epsilon, t)=\mathcal{I}^{\rm (inel)}(\epsilon)+\mathcal{I}^{\mathrm{(e-ph)}}(\epsilon)+\mathcal{I}^{(e-e)}(\epsilon),
\end{equation}
where $\delta$ is the mean level spacing in the grain. The left hand side of this equation describes the 
change of the electron distribution function with time $t$.
The right hand side is the sum of collision integrals with $\mathcal I^{\rm (inel)}(\epsilon)$ being the scattering integral due to
cotunneling processes, while $\mathcal I^{\mathrm{(e-ph)}}(\epsilon)$ and $\mathcal I^{(e-e)}(\epsilon)$ being the electron-phonon and electron-electron scattering integrals, respectively. We write the scattering integral $\mathcal I^{\rm (inel)}(\epsilon)$ explicitly in Appendix~\ref{Ap1}.

We emphasize that only \textit{inelastic} cotunneling contributes to the scattering integral $\mathcal{I}^{\rm (inel)}(\epsilon)$ in Eq.~\eqref{keq}. It satisfies the particle conservation law
\begin{gather}\label{Iparticle}
\int \,\mathcal{I}^{\rm (inel)}(\epsilon)d\epsilon\equiv 0.
\end{gather}
This property does not contradict the current flow through the junction since the grain  can be considered as a \textit{virtual} haven for tunneling electrons during the cotunneling processes. However, the scattering integral $\mathcal I^{\rm (inel)}(\epsilon)$ does not conserve energy, since each inelastic cotunneling process leaves an excited electron-hole pair behind in the grain. The heat dissipation rate into the grain is
\begin{gather}\label{dQ}
  \dot Q=\int\epsilon \,\mathcal{I}^{\rm (inel)}(\epsilon)d\epsilon.
\end{gather}
Here the energy $\epsilon$ in the electron grain distribution function $f^{(g)}(\epsilon)$ is counted from the local electrochemical potential.

\subsection{Local equilibrium approximation}

To solve the kinetic equation~\eqref{keq} we use a {\em local equilibrium approximation}.
This approximation is valid because for small grains the electron-electron (Coulomb) interaction is strong. Therefore the effective electron-electron scattering time, $\tau_{e-e}$, corresponding to the scattering integral $\mathcal{I}^{(e-e)}(\epsilon)$ in Eq.~\eqref{keq} is shorter than the inelastic cotunneling scattering time, $\tau_{\rm inel}$ and the electron-phonon scattering time, $\tau_{\rm e-ph}$,~\cite{glatz+prb11}
\begin{equation}
\label{inequlity}
\tau_{e-e} \ll {\rm min}
(\tau_{\rm inel}, \, \tau_{\rm e-ph}).
\end{equation}
In addition, the scattering integral $\mathcal{I}^{(e-e)}(\epsilon)$ in Eq.~\eqref{keq} conserves the energy and the particle number. Therefore we can find the solution of the kinetic equation~\eqref{keq} using the local equilibrium approximation
\begin{equation}
\label{local}
f^{(g)}(\epsilon)\approx f_F(\epsilon,T_g).
\end{equation}
This expression substituted into the scattering integral $\mathcal{I}^{(e-e)}(\epsilon)$ in Eq.~\eqref{keq} lets it vanish.
Using the particle conservation law, Eq.~\eqref{Iparticle}, the correction to the effective electro-chemical potential in the grain in the local equilibrium approximation can be neglected. Therefore in this approximation there is only one unknown parameter, the grain temperature $T_g$.

\subsection{Heat balance equation}

The heat rate between electrons and phonons, 
\begin{equation}
\dot q = \int \epsilon \mathcal{I}^{\mathrm{(e-ph)}}(\epsilon)\, d\epsilon,
\end{equation}
in the local equilibrium approximation, Eq.~\eqref{local}, is given by Eq.~(\ref{ephq}).
Therefore the problem of finding the grain temperature $T_g$ in the local equalibrium approximation, Eq.~\eqref{local}, reduces to the solution of the heat balance equation that follows from Eq.~\eqref{keq}
\begin{gather}\label{hb}
  \dot q(T_g,T_{\rm ph}) = \dot Q(T_g,T_{\ssL},T_{\ssR},V).
\end{gather}

Solving Eq.~\eqref{hb} we find the grain temperature $T_g(T_{\ssL},T_{\ssR},V,T_{\rm ph})$ as a function of lead temperatures $T_{\ssL}$, $T_{\ssR}$, the bias voltage $V$, and the phonon temperature $T_{\rm ph}$. Using the result for grain temperature $T_g(T_{\ssL},T_{\ssR},V,T_{\rm ph})$ we find the current-voltage characteristics in Eq.~\eqref{current-voltage}. Below we proceed with this program focusing on the analytical solution of Eqs.~\eqref{current-voltage} and~\eqref{hb}.

\section{Bosonic representation of charge and heat rates}

Here we consider the expressions for the inelastic cotunneling rate ${\Gamma}^{\rm(inel)}$ in Eq.~\eqref{inelastic_rate} and for the heat dissipation rate $\dot Q$ in Eq.~\eqref{dQ} in more details.

For small grains the electrostatic energies, $\Delta F_{\ssL},^+(n)$ and $\Delta F_{\ssR},^-(n)$ in Eq.~\eqref{inelastic_rate}, are much larger than all other characteristic energy scales in the problem including temperature and voltage. Therefore we can neglect in the denominators of Eqs.~\eqref{inelastic_rate},~\eqref{Icot}, and~\eqref{inelastic_Qrate} the energy $\epsilon$ and $\epsilon_i$, $i=1,\ldots,4$. This limit allows for an analytical solution of the heat balance equation~\eqref{hb} in order to calculate the grain temperature $T_g$. In addition, we show that it is more convenient to rewrite all transport characteristics in terms of electron-hole (dipole) excitations to find the current-voltage characteristics in Eq.~\eqref{current-voltage}.

\begin{figure}[h]
  \centering
  \includegraphics[width=25mm]{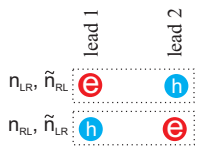}\\
  \caption{The form-factor $n^{\LR}_{\omega}$ ($\tilde n^{\RL}_{\omega}$)  can be interpreted as the distribution function of electron-hole pairs where the electron sits at the first lead while the hole on the second lead. The form-factor $n^{\RL}_{\omega}$ ($\tilde n^{\LR}_{\omega}$) corresponds to the opposite situation.}\label{figehn}
\end{figure}

First, we perform the substitution  $\epsilon_1\to\epsilon_1-eV/2$ and $\epsilon_4\to\epsilon_1+eV/2$ in Eq.~\eqref{inelastic_rate}, which moves the voltage dependence into the distribution functions and in particular implies that $f_{{\ssL}({\ssR})}(\epsilon)=f_F(\epsilon\mp eV/2,T_{{\ssL}({\ssR})})$. After the energy-integration, we obtain the following expression for the inelastic cotunneling rate (see Appendix~\ref{Ap2})\cite{units}
\begin{subequations} 
\label{rates}
\begin{multline}\label{Ginb}
    \overrightarrow{\Gamma}^{\rm(inel)}=  \frac{1}{2\pi^3} \frac{R_q^2}{R_{\ssL}R_{\ssR}} \frac{1}{\hbar} 
    \times
    \\
    \int_0^\infty d\omega\,\frac{\omega^2}{E_c^2} \left\{n_{\omega}^{\LR} [N_\omega+1] +[1+\tilde n_{\omega}^{\LR}] N_\omega\right\},
\end{multline}
where we used the notation
\begin{gather}\label{Ec}
  \frac1{E_c}\equiv\frac{1}{\Delta F_L^+(n)}+\frac{1}{\Delta F_R^-(n)}.
\end{gather}
\end{subequations}
For the heat dissipation rate in the grain we obtain, using Eqs.~\eqref{qrates}-\eqref{inelastic_Qrate}:
\begin{multline}\label{Qb}
    \dot Q = \frac{1}{\pi^3} \frac{R_q^2}{R_{\ssL}R_{\ssR}} \frac{1}{\hbar E_c^2} \times
    \\
    \int_{0}^\infty \omega^3 d\omega\, \left\{n_{\omega} [N_{\omega}+1] -[1+n_{\omega}] N_{\omega}\right\}.
\end{multline}
Here we introduce the following form-factors, see Fig.~\ref{figehn}, that describe the nonequilibrium electron-hole pairs~\cite{footnote}
\begin{subequations}\label{eqnlrall}
\begin{gather}\label{eqnlr}
    n^{\LR}_{\omega}=\frac 1\omega\int_{-\infty}^\infty d\epsilon f^{(L)}(\epsilon_+) \left[1-f^{(R)}(\epsilon_-)\right],
    \\
    n^{\RL}_{\omega}=\frac 1\omega\int_{-\infty}^\infty d\epsilon f^{(R)}(\epsilon_+) \left[1-f^{(L)}(\epsilon_-)\right],
    \\\label{tn}
    1+\tilde n^{\LR}_{\omega}\equiv -n^{\LR}_{-\omega},\\ 1 + \tilde n^{\RL}_{\omega}\equiv -n^{\RL}_{-\omega},
\end{gather}
\end{subequations}
where $\epsilon_{\pm} = \epsilon \pm \omega/2$. General identities coupling the bosonic form-factors with and without tilde [see proof in Appendix~\ref{apboson}] are given by:
\begin{gather}\label{nidentity1}
                 \tilde n^{\LR}_\omega-n^{\RL}_\omega= \frac{eV}{\omega},
                 \\\label{nidentity2}
                 \tilde n^{\RL}_\omega-n^{\LR}_\omega=-\frac{eV}{\omega}.
               \end{gather}

Finally we define the form-factors $n_{\omega}$ and $N_{\omega}$ in Eqs.~\eqref{rates}
\begin{subequations}\label{eqnall}
\begin{gather}\label{eqn}
    n_{\omega} =\frac 12 \left\{n^{\LR}_{\omega} + n^{\RL}_{\omega}\right\},
    \\\label{eqN}
    N_{\omega} =\frac 1\omega\int_{-\infty}^\infty d\epsilon f^{(g)}(\epsilon_+) \left[1-f^{(g)}(\epsilon_-)\right].
\end{gather}
\end{subequations}

The functions $n_{\omega}$ and $N_{\omega}$ satisfy the basic property of the Bose-function $N_B(\omega,T)=1/[\exp(\omega/T)-1]$, $N_B(-\omega,T)=-[1+N_B(\omega,T)]$, e.g., $n_{-\omega} = -[1 + n_{\omega}]$. Below we refer to the distribution functions $n_{\omega}$ and $N_{\omega}$ as to the electron-hole distribution functions in the leads and in the grain, respectively.

\subsection{Physical interpretation of equations~(\ref{rates}) - (\ref{eqnall}).}

The effective electron-hole distribution function $n_{\omega}$ in Eq.~\eqref{eqn} has a direct physical meaning. 
It describes the concentration of electron-hole pairs where the electron is localized in the left and the hole in the right lead or vice versa~\cite{ch0,ch01,ch1,glatz+prb12}.

The form-factor $n^{\LR}_{\omega}$ in Eq.~\eqref{eqnlr} is more specific than $n_{\omega}$. 
It describes the concentration of ``polarized'' electron-hole pairs with
electron being sitting strictly in the left lead while the hole being occupying the right lead, see Fig.~\ref{figehn}. Therefore the annihilation of such an electron-hole pair (or the creation of the ``RL'' electron-hole pair) leads to charge transfer from the left to the right lead.

The expression, $n_{\omega}^{\LR}[N_\omega+1]$, in the r.~h.~s. of, Eq.~\eqref{Ginb}, is proportional to the probability of annihilation of an electron in the left lead and a hole in the right lead and the creation of an electron-hole pair localized in the grain. The physical realization of such a process leads to a charge transfer from the left to the right lead. The product, $[1+\tilde n_{\omega}^{\RL}]N_\omega$,  is proportional to the probability of the creation of a hole in the left lead and an electron in the right lead and the annihilation of an electron-hole pair localized in the grain. The realization of this process also leads to a charge transfer from the left to the right lead.

The expression, $n_{\omega}[N_\omega+1]$, in the r.~h.~s. of Eq.~\eqref{Qb} is proportional to the probability for the annihilation of an electron-hole pair in the leads (electron in the left lead and a hole in the right lead and vice versa) and the creation of an electron-hole pair localized in the grain. This process increases the energy in the grain, but does not necessarily imply a charge transfer between the leads.  Finally, the product $[1+n_{\omega}] N_{\omega}$ corresponds to the reverse process that cools the grain.

\subsection{Cotunneling current-voltage characteristics}

Using Eq.~\eqref{Ginb} we can write the inelastic cotunneling current in the following form
\begin{multline}\label{IinelN}
    I^{\rm (inel)}= \frac{1}{2\pi^3} \frac{R_q^2}{R_{\ssL}R_{\ssR}} \frac{e}{\hbar} 
    \int_0^\infty d\omega\,\frac{\omega^2}{E_c} \times
    \\
    \left\{[n_{\omega}^{\LR}- n_{\omega}^{\RL}][N_\omega+1] +[\tilde n_{\omega}^{\LR}-\tilde n_{\omega}^{\RL}] N_\omega\right\}.
\end{multline}
Identities~\eqref{nidentity1}-\eqref{nidentity2} allow us to simplify  Eq.~\eqref{IinelN} and separate the grain and the lead degrees of freedom. So we get for the integral in Eq.\eqref{IinelN}
\begin{gather}\label{IinelNf}
     I^{\rm (inel)}= \frac{1}{2\pi^3} \frac{R_q^2}{R_{\ssL}R_{\ssR}} \frac{e}{\hbar}
\int\limits_0^\infty d\omega\,\frac{\omega^2}{E_c^2} \left\{[n_{\omega}^{\LR}- n_{\omega}^{\RL}] +\frac{2eV}\omega N_\omega\right\}.
\end{gather}
The integration over the energy $\omega$ in Eq.~\eqref{IinelNf} leads to the final result for the inelastic
cotunneling current given in Eq.~(\ref{current-voltage}), where we used Eqs.~\eqref{N}-\eqref{TN} for the integration of the  $n_{\omega}^{\LR,\RL}$ parts of the integral in Eq.~\eqref{IinelNf}.

\subsection{Evaluation of the heat dissipation rate $\dot Q$ in the grain}

Formally the heat dissipation rate into the grain $\dot Q$ in Eqs.~\eqref{dQ} and \eqref{Qb} can be obtained using the effective kinetic equation for electron-hole pairs
\begin{subequations}
\begin{gather}\label{dotN}
 \frac{d}{d(\hbar^{-1} E_ct)} N_\omega=\frac{n_{\omega}(N_{\omega}+1) -N_{\omega}(1+n_{\omega})}{\tau^{\rm (inel)}(\omega)},
\end{gather}
where the relaxation rate for electron-hole pairs in the grain is mediated by their interaction with cotunneling electrons
\begin{gather}
\label{tau}
  \frac{1}{\tau^{\rm (inel)}(\omega)}\equiv \frac{1}{\pi^3} \frac{R_q^2}{R_{\ssL}R_{\ssR}} \frac{\omega^2}{E_c^2}.
\end{gather}
\end{subequations}
Equation~\eqref{tau} means that the relaxation rate of electron-hole pairs in the grain in the cotunneling channel is proportional to the second power of temperature at small temperatures, $1/\tau^{\rm (inel)}(\omega) \sim T^2$.

For very low temperatures, where electron-electron and electron-phonon interactions are frozen out and scattering due to cotunneling is the leading scattering mechanism we obtain $N_\omega = n_\omega$ using Eq.~\eqref{dotN}.

However, for a small metallic grain the electron-electron interaction is the main scattering mechanism. It drives the electron-hole distribution function, $N_{\omega}$, to the local-equilibrium form, $N_{\omega}\approx N_B(\omega,T_g)$. [The last statement follows from the fact that in the local equilibrium approximation the distribution function $f^{(g)}\approx f_F(\epsilon,T_g)$ and therefore according to Eq.~\eqref{eqN}, $N_{\omega}\approx N_B(\omega,T_g)$.] In this case the effective grain temperature $T_g$ is determined by the heat-balance equation~\eqref{hb}.

To evaluate the heat dissipation rate in the grain, $\dot Q$ in Eq.~\eqref{Qb}, we need to know the integral
\begin{gather}\label{integral}
\dot{Q} = \frac{1}{\pi^3} \frac{R_q^2}{R_{\ssL}R_{\ssR}} \frac{1}{\hbar E_c^2}
\int_{0}^\infty \omega^3  \left(n_{\omega}  -N_{\omega}\right) \,d\omega.
\end{gather}
Using Eq.~\eqref{eqn} and Eqs.~\eqref{N}-\eqref{TN} (see  Appendix~\ref{Ap2}) for the first term in this integral we obtain  $\int_{0}^\infty \omega^3  n_{\omega}\,d\omega= T_{\rm eff}^4\pi^4/15$, where
\begin{gather}\label{Teff}
T_{\rm eff}^4=T_{\rm m}^4+\frac5{2\pi^2}T_{\rm m}^2(eV)^2+\frac{5}{8\pi^4}(eV)^4,
\end{gather}
being the effective temperature.
The second contribution to the integral in Eq.~\eqref{integral} leads accordingly to
$\int_{0}^\infty \omega^3 N_{\omega}\,d\omega = T_{g}^4\pi^4/15$.
Thus we obtain our main result, Eq.~\eqref{Q}, with parameter $a$ given by the expression
\begin{gather}
\label{a1}
  a = \frac{\pi}{15} \frac{R_q^2}{R_{\ssL}R_{\ssR}} \frac{1}{\hbar E_c^2}.
\end{gather}
Here we stress that the effective temperature $T_{\rm eff}$ in Eq.~\eqref{Teff}, follows the limit,
$\lim_{\omega\to 0}\omega n_\omega=\frac{eV}{2} \coth\frac{eV}{2T_{\rm m}}$, only for small bias voltages, $eV\lesssim T_{\rm m}$.

\section{Discussion}

\subsection{Quasi-equilibrium limit of the current voltage charateristics}

For equal lead and grain temperatures, $T_{\ssL}=T_{\ssR}=T_g$, we reproduce the known result for the cotunneling current, Refs.~[\onlinecite{Beloborodov07,Averin92}]. However, in the general case, the inelastic current $I^{\rm (inel)}$ in Eqs.~\eqref{IinelN} 
and (\ref{IV1}) strongly depends on both the grain and the lead temperatures.

\subsection{Heat dissipation rate and total power}

The total power of the junction, given by the product of current and bias voltage, $I\,V$ and the heat dissipated into the grain, $\dot Q(T_g,T_1,T_2,V)$, are two different quantities.
Indeed, the heat $\dot Q(T_g,T_{\ssL},T_{\ssR},V)$ is finite even for zero bias voltage, $V=0$, if the lead temperatures are different from the grain temperature, $T_{{\ssL},{\ssR}}\neq T_g$, while the total power is zero, $I\,V=0$, in this case.

For finite voltage, $V\neq0$, and equal lead temperatures, $T_{\ssL} = T_{\ssR}$, the heat dissipation rate $\dot Q$ can be identified with the Joule heat released {\it in the grain}. There is another part of the Joule heat, $\dot Q_{\rm leads}$, corresponding to the energy equilibration of nonequilibrium electrons in the bulk of the lead after they co-tunnel from the other lead. Therefore the Joule heat is the sum of both contributions, $I\,V=\dot Q_{\rm leads}+\dot Q$.

It is interesting to compare the heat dissipation rate in the grain $\dot Q$ with the Joule heat, $I\, V$.  Using Eq.~\eqref{Q} for the heat dissipation rate in the grain $\dot Q$ and an explicit expression  for inelastic part of cotunneling current Eq.~\eqref{IV1} we obtain
 \begin{gather} \label{ratio}
 \frac{\dot Q}{I^{\rm (inel)}\, V}= \frac{4\pi^4}5\frac{T_{\rm eff}^4-T_g^4}{\left\{(eV)^2+2\pi^2(T_{\rm m}^2+T_g^2)\right\}(eV)^2}.
\end{gather}

First, we consider the case of equal temperatures, $T=T_{\ssL}=T_{\ssR}=T_g$. In this case we obtain
\begin{gather}
\label{ratio2}
 \frac{\dot Q}{I^{\rm (inel)}\, V}= \frac{1}{2}.
\end{gather}
Equation~\eqref{ratio2} has transparent physical meaning - half of the energy in the cotunneling process is spent to generate  electron-hole pairs in the grain and the other half to inject the nonequilibrium electrons into
the leads that finally equilibrate in the bulk of the leads.
Remarkably,  the ratio $\dot Q/(I^{\rm (inel)}\, V)$ is universal and does not depend on temperature $T$ and voltage $V$.

\subsection{Transport in semiconducting nanojunction and spectrum of bosonic modes in the grain}

Calculating the current and the heat we implicitly assume that the grain is metallic meaning that electron-hole pairs with an arbitrary small energy $\omega$ can be created during the inelastic cotunneling process. 
However, if the grain is made of a semiconducting material the spectrum of electron-hole pairs would have a gap $\Delta$. 
In this case the lower limit of integration over the energy $\omega$ in Eq.~\eqref{IinelN} would be $\Delta$ and a smooth weight function $\rho(\omega)$  would appear  under the integral renormalizing the interaction vertex of electron-hole pairs with the cotunneling electrons.  
Therefore in this case the current would be exponentially suppressed for voltages $V$ and temperatures $T=T_{\ssL}=T_{\ssR}=T_g$ smaller than the gap $\Delta$, $I^{\rm (inel)}(V)\sim V\exp(-\Delta/T)$ .
This shows that the current-voltage characteristics is very sensitive to the spectrum of the grain.

\section{Conclusions}

We studied the charge transport and the heat transfer through a small metallic grain weakly coupled to two metallic leads.
We focused on the cotunneling regime out-of-equilibrium, when the bias voltage and the temperature gradient between the leads strongly drive electron and phonon degrees of freedom in the granule that in turn have a strong feedback on transport through the grain.  
We derived and solved the coupled kinetic equations for electron and phonon degrees of freedom in the grain, found the heat fluxes between cotunneling electrons, bosonic electron-hole excitations in the grain, and phonons, and selfconsistently obtained the current-voltage characteristics. We demonstrated that the transport in the nanojunction is very sensitive to the spectrum of the bosonic modes in the grain.

\begin{acknowledgments}
A.~G. was supported by the U.S. Department of Energy Office of Science under the Contract No. DE-AC02-06CH11357.
I.~B. was supported by NSF Grant DMR 1158666.
\end{acknowledgments}

\appendix

\section{Scattering integral due to cotunneling\label{Ap1}}

Here we present the scattering integral $\mathcal I^{\rm (inel)}(\epsilon)$ in Eq.~\eqref{keq} induced by cotunneling processes
\begin{widetext}
\begin{multline}\label{Icot}
 \mathcal{I}^{\rm(inel)}(\epsilon)= \frac{1}{2\pi^3} \frac{R_q^2}{R_{\ssL}R_{\ssR}} \frac{1}{\hbar}\int  d\epsilon_1d\epsilon_3d\epsilon_4f^{(L)}(\epsilon_1)[1-f^{(g)}(\epsilon)]f^{(g)}(\epsilon_3)[1-f^{(R)}(\epsilon_4)]\times
 \\
 \left\{\frac{1}{\epsilon-\epsilon_1+\Delta F_L^+(n)}+\frac{1}{\epsilon_4-\epsilon_3+\Delta F_R^-(n)}\right\}^2
 \delta(eV+\epsilon_1-\epsilon+\epsilon_3-\epsilon_4)-
 \\
\frac{1}{2\pi^3} \frac{R_q^2}{R_{\ssL}R_{\ssR}} \frac{1}{\hbar}
 \int  d\epsilon_1d\epsilon_2d\epsilon_4f^{(L)}(\epsilon_1)[1-f^{(g)}(\epsilon_2)]f^{(g)}(\epsilon)[1-f^{(R)}(\epsilon_4)]\times
 \\
 \left\{\frac{1}{\epsilon_2-\epsilon_1+\Delta F_L^+(n)}+\frac{1}{\epsilon_4-\epsilon+\Delta F_R^-(n)}\right\}^2
 \delta(eV+\epsilon_1-\epsilon_2+\epsilon-\epsilon_4)+\left(L\rightleftarrows R, \,V\to-V\right),
\end{multline}
where the last term in the brackets results from the contribution of the electron cotunneling rate
$\overleftarrow{\Gamma}^{\rm(inel)}$ from the right to the left leads. Elastic
cotunneling does not contribute to the scattering integral.

Using Eq.~\eqref{Icot} we rewrite the heat flux dissipated into the grain in terms of the heat rates similar to Eq.~\eqref{inelastic_rate}:
\begin{gather}\label{qrates}
  \dot Q=\overrightarrow{\Gamma}_Q+\overleftarrow{\Gamma}_Q,
\end{gather}
where the heat rate from the left to the right lead is
\begin{multline}\label{inelastic_Qrate}
 \overrightarrow{\Gamma}_Q = \frac{1}{2\pi^3} \frac{R_q^2}{R_{\ssL}R_{\ssR}} \frac{1}{\hbar} \int d\epsilon_1d\epsilon_2d\epsilon_3d\epsilon_4(\epsilon_2-\epsilon_3)f^{(L)}(\epsilon_1)[1-f^{(g)}(\epsilon_2)]f^{(g)}(\epsilon_3)[1-f^{(R)}(\epsilon_4)]\times
 \\
 \left\{\frac{1}{\epsilon_2-\epsilon_1+\Delta F_L^+(n)}+\frac{1}{\epsilon_4-\epsilon_3+\Delta F_R^-(n)}\right\}^2
 \delta(eV+\epsilon_1-\epsilon_2+\epsilon_3-\epsilon_4).
\end{multline}
\end{widetext}
The plus sign in Eq.~\eqref{qrates} implies that the particular direction of the cotunneling process, from the left to the 
right or from the right to the left, is not important for heating or freezing of the grain.

\section{Bosonic representation of charge and heat rates \label{Ap2}}

\subsection{Inelastic scattering rate}

Integrating the expression for the inelastic cotunneling rate $\overrightarrow{\Gamma}^{\rm(inel)}$ in Eq.~\eqref{inelastic_rate} over the $E=(\epsilon_2+\epsilon_3)/2$, $E'=(\epsilon_1+\epsilon_4)/2$ and introducing new variables $\omega=\epsilon_2-\epsilon_3$ and $\omega'=\epsilon_1-\epsilon_4$, we obtain
\begin{multline}\label{gg}
    \int_{\epsilon_1,\ldots,\epsilon_4} f^{(L)}_{\epsilon_1}[1-f^{(R)}_{\epsilon_4}] [1-f_{\epsilon_2}^{(g)}]f^{(g)}_{\epsilon_3}\delta_{\epsilon_1-\epsilon_2+\epsilon_3-\epsilon_4}=
    \\
    \int d\omega d\omega' dEdE' f^{(L)}_{E'_+}[1-f^{(R)}_{E'_-}] [1-f^{(g)}_{E_+}]f^{(g)}_{E_-}\delta(\omega-\omega')=
    \\
    =\int_{-\infty}^\infty d\omega \omega^2 n_{\omega}^{\LR} [1+N_\omega],
\end{multline}
where $E_{\pm}=E\pm\omega/2$ and $E'_{\pm}=E'\pm\omega'/2$. The form factors $n^{\LR}_{\omega}$ and $N_{\omega}$ in Eq.~\eqref{gg} can be interpreted as the effective distributions of electron-hole pairs, see discussions below Eqs.~\eqref{eqnlr} and~\eqref{eqn}. This interpretation is possible for positive frequencies $\omega > 0$. Therefore it is convenient to
transform the integral in the last line of Eq.~\eqref{gg} and find
\begin{multline}\label{ggg}
    \int_{0}^\infty d\omega \omega^2 \left\{n_{\omega}^{\LR} [1+N_\omega]+n_{-\omega}^{\LR} [1+N_{-\omega}]\right\}\equiv
    \\
    =\int_{0}^\infty d\omega \omega^2 \left\{n_{\omega}^{\LR} [1+N_\omega]+[1+\tilde n_{\omega}^{\LR}] N_{\omega}\right\}.
\end{multline}
This result immediately leads to Eq.~\eqref{Ginb} for the inelastic scattering rate $\overrightarrow{\Gamma}^{\rm(inel)}$
given in the text.

\subsection{Bosonic form-factors\label{apboson}}

To prove identities, Eqs.~\eqref{nidentity1}-\eqref{nidentity2},  we rewrite Eq.~\eqref{tn},
\begin{multline}
  1+\tilde n^{\LR}_{\omega}=\frac 1\omega\int_{-\infty}^\infty d\epsilon f^{(L)}(\epsilon_-) \left[1-f^{(R)}(\epsilon_+)\right]=
  \\
  n^{\RL}_\omega-\frac 1\omega\int_{-\infty}^\infty d\epsilon \left[f^{(R)}(\epsilon_+)-f^{(L)}(\epsilon_-)\right].
\end{multline}
Taking the distribution functions $f^{(L)}(\epsilon,T_1)=f_F(\epsilon-eV/2,T_1)$ and $f^{(R)}(\epsilon,T_2)=f_F(\epsilon+eV/2,T_2)$ and using, $\int_{-\infty}^\infty[f_F(\epsilon,T_1)-f_F(\epsilon,T_2)]d\epsilon=0$, we find for the last integral:
\begin{multline}
  \frac 1\omega\int_{-\infty}^\infty d\epsilon \left[f^{(R)}(\epsilon_+)-f^{(L)}(\epsilon_-)\right]=\frac{\omega+eV}\omega\times
  \\
\int_{-\infty}^\infty d\epsilon\frac{\partial}{\partial\epsilon} \frac12\left[f^{(R)}(\epsilon)+f^{(L)}(\epsilon)\right]=-1-\frac {eV}{\omega}.
\end{multline}
This proves the validity of Eqs.~\eqref{nidentity1}-\eqref{nidentity2}.

Using the distribution functions $f^{(L)}(\epsilon,T_{\ssL})=f_F(\epsilon-eV/2,T)$ and $f^{(R)}(\epsilon,T_{\ssR})=f_F(\epsilon+eV/2,T)$ in Eq.~\eqref{inelastic_Qrate},
with $T$ being the temperature of both leads, we find the explicit form of the form-factors\cite{footnote}
\begin{align}\label{N}
    n^{\LR}_{\omega}&=\frac {\omega-eV}\omega N_B(\omega-eV,T),
    \\
    \tilde n^{\LR}_{\omega}&=\frac{eV}{\omega} +\frac {\omega+eV}\omega N_B(\omega+eV,T),
    \\\label{NRL}
    n^{\RL}_{\omega}&=\frac {\omega+eV}\omega N_B(\omega+eV,T),
    \\\label{TN}
    \tilde n^{\RL}_{\omega}&=-\frac {eV}{\omega} +\frac {\omega-eV}\omega N_B(\omega-eV,T).
\end{align}

Equations~\eqref{N}-\eqref{TN} are exact for equal lead temperatures only, $T=T_{\ssL} = T_{\ssR}$.
However, if the lead temperatures are different $T_{\ssL}\neq T_{\ssR}$ then Eqs.~\eqref{N}-\eqref{TN} give still a good approximation of the form-factors if we use the substitution, $T\to T_{\rm m}=\sqrt{\frac{T_{\ssL}^2+T_{\ssR}^2}2}$.
This is a good approximation for $T_{{\ssL},{\ssR}}\gtrsim eV$ and for $T_{{\ssL},{\ssR}}<eV$ when $T_{\ssR}\sim T_{\ssL}$.
Furthermore, it reproduces $\lim_{\omega\to 0} \omega n^{\LR,\RL}_\omega=\frac{eV}{2}\coth\frac{eV}{2T_{\mathrm{m}}}$.

The integrals of form $\int_{0}^\infty \omega^a n_{\omega} d\omega$, $a=2,3,\ldots$ can be calculated analytically using  Eqs.~\eqref{eqn},~\eqref{N} and \eqref{NRL}, and expressed through the polylogarithms\cite{integrals}. So, for instance,
\begin{multline}
  \int_{0}^\infty \omega^3 n_{\omega}\,d\omega =
 \\
\left\{ \left[\Li_{3}(e^{-v}) - \Li_{3}(e^{v})\right]v +
 3 \left[\Li_{4}(e^{-v}) + \Li_{4}(e^{v})\right]\right\}=
 \\
 -\pi^4 \Brnli_4\left(-\frac{v}{2\pi i}\right)-\frac{4\pi^3}3v \Brnli_3\left(-\frac{v}{2\pi i}\right)=
 \\
 \frac{(\pi T)^4}{15}\left\{1+\frac5{2\pi^2}v^2+\frac{5}{8\pi^4}v^4\right\},
\end{multline}
where we use the notation $v=eV/T$.


\begin{thebibliography}{99}
\bibitem{units}
We use the units $k_B=1$ and $\omega$ variables refer to energy quantities.
\bibitem{footnote}
If we defined the formfactors differently: $n^{\LR}_{\omega}=\frac 1{\omega+V}\int_{-\infty}^\infty d\epsilon f^{(L)}(\epsilon_+) \left[1-f^{(R)}(\epsilon_-)\right]$, and similarly, $n^{\RL}_{\omega}=\frac 1{\omega-V}\int_{-\infty}^\infty d\epsilon f^{(R)}(\epsilon_+) \left[1-f^{(L)}(\epsilon_-)\right]$, then we would get: $\tilde n^{\LR}_{\omega}\equiv n^{\RL}_{\omega}$ and $n^{\LR}_{\omega}\equiv \tilde n^{\RL}_{\omega}$. However then the factors $\omega\pm V$ would appear explicitly in the expressions for $I$ and $\dot Q$ renormalizing the boson-boson interaction vertices.
\bibitem{integrals}
More generally, integrals of this type can be expressed in terms of polylogarithms, $\Li_n(x)$, using $\int_0^\infty\, x^n/\left(e^{x+v}-1\right)dx=n!\Li_{1+n}(e^{-v})$.
The $\Li_n$-functions can be rewritten in terms of Bernoulli polynomials, $\Brnli_n$, using the identity, $\Li_n(z)+(-1)^n\Li_n(1/z) =-\frac{(2\pi i)^n}{n!}\Brnli_n\left[\frac12+\frac{\ln(-z)}{2\pi i}\right]$. The latter identity gives the result for $n_{\omega}$ directly.
\end{thebibliography}
\end{document}